\documentclass{emulateapj}

\usepackage{natbib}
\usepackage{lscape}
\usepackage{apjfonts}
\slugcomment{Accepted by ApJ 2009, October}

\shorttitle{The Eclipsing Neutron Star 4U~2129+47}

\shortauthors{Lin, Nowak, \& Chakrabarty}

\begin{document}

\title{A Further Drop into Quiescence by the Eclipsing Neutron Star 4U~2129+47}

\author{Jinrong Lin\altaffilmark{1}, Michael A. Nowak\altaffilmark{1},
        Deepto Chakrabarty\altaffilmark{1}}
\altaffiltext{1}{Massachusetts Institute of Technology, Kavli
  Institute for Astrophysics and Space Research, Cambridge, MA 02139, USA;
 jinrongl@mit.edu; mnowak,deepto@space.mit.edu}

\begin{abstract}
The low mass X-ray binary 4U~2129+47 was discovered during a
previous X-ray outburst phase and was classified as an accretion
disk corona source. A $1\%$ delay between two mid-eclipse epochs
measured $\sim 22$ days apart was reported from two {\it
XMM-Newton} observations taken in 2005, providing support to the
previous suggestion that 4U~2129+47 might be in a hierarchical
triple system. In this work we present timing and spectral
analysis of three recent {\it XMM-Newton} observations of
4U~2129+47, carried out between November 2007 and January 2008. We
found that absent the two 2005 {\it XMM-Newton} observations, all
other observations are consistent with a linear ephemeris with a
constant period of $18~857.63$\,s; however, we confirm the time
delay reported for the two 2005 {\it XMM-Newton} observations.
Compared to a {\it Chandra} observation taken in 2000, these new
observations also confirm the disappearance of the sinusoidal
modulation of the lightcurve as reported from two 2005 {\it
XMM-Newton} observations. We further show that, compared to the
{\it Chandra} observation, all of the {\it XMM-Newton}
observations have 40\% lower 0.5--2\,keV absorbed fluxes, and the
most recent {\it XMM-Newton} observations have a combined
2--6\,keV flux that is nearly 80\% lower. Taken as a whole, the
timing results support the hypothesis that the system is in a
hierarchical triple system (with a third body period of at least
175 days). The spectral results raise the question of whether the
drop in soft X-ray flux is solely attributable to the loss of the
hard X-ray tail (which might be related to the loss of sinusoidal
orbital modulation), or is indicative of further cooling of the
quiescent neutron star after cessation of residual, low-level
accretion.

\end{abstract}


\section{Introduction}

4U 2129+47 was discovered to be one of the Accretion Disk Corona (ADC)
sources \citep{for78}, which are believed to be near edge-on accreting
systems since they have shown binary orbital modulation via broad,
partial V-shape X-ray eclipses (\citealt{tho79,mcc82}, hereafter MC82,
\citealt{wh82}); the eclipse width was $\approx 0.2$ in phase and
$\approx 75\%$ of the X-rays were occulted at the eclipse midpoint.
The origins of ADCs are not fully understood, but they are typically
associated with high accretion-rate systems, wherein we are only
observing the small fraction of the system luminosity that is
scattered into our line of sight. For the prototypical ADC
X1822$-$371, models suggest that it is accreting at a near Eddington
rate, with the radiation scattered into our line of sight having an
equivalent isotropic luminosity on the order of 1\% of the total
luminosity (see \citealt{par00,hei01,cot01}, and references therein).

In the early 1980s, observations of 4U 2129+47 showed that both its
X-ray and optical lightcurves were modulated over a $5.24$~h period
\citep{tho79,ulm80,mcc82,wh82}. The discovery of a type-I X-ray burst
led to the classification of 4U 2129+47 as a neutron star (NS) low
mass X-ray binary (LMXB) system \citep{gar87}, and the companion was
suggested to be a late K or M spectral type star of $\sim 0.6
M_{\bigodot}$. The source distance was estimated to be $\sim
1$--2\,kpc \citep{hor86}. Assuming isotropic emission, the X-ray
luminosity would have then corresponded to $\sim
5\times10^{34}$~ergs$^{-1}$. Even if the luminosity were 100 times
larger, the luminosity would have been somewhat smaller than expected
for an ADC.

Since 1983, 4U 2129+47 has been in a quiescent state. Optical
observations show a flat lightcurve without any evidence for orbital
modulation between the years 1983 and 1987. Additionally, instead of
an expected M- or K-type companion, the observed spectrum was
compatible with a late type F8 IV star \citep{kal88,che89}. The
refined X-ray source position as determined by {\it Chandra} turned
out to be coincident with the F star to within $0.1^{''}$
(\citealt{now02}, hereafter N02). The probability of a chance
superposition is less than $10^{-3}$ (see also \citealt{both08}),
therefore, the hypothesis of a foreground star is unlikely. A $\sim
40$\,km~s$^{-1}$ shift in the mean radial velocity was derived from
the F star spectrum, providing evidence for a dynamical interaction
between the F star and 4U 2129+47 \citep{cow90,both08}. The system was
therefore suggested to be a hierarchical triple system, in which the F
star is in a month-long orbit around the binary \citep{gar89}. This
hypothesis is tentatively confirmed with two {\it XMM-Newton}
observations separated by 22 days that showed deviations from a simple
orbital ephemeris (\citealt{boz07},hereafter B07). Assuming that the F
star is part of the system, the source distance was revised to
$\sim6.3$~kpc \citep{cow90}. This would imply that the peak luminosity
observed prior to 1983 (assuming that the observed flux was $\sim1$\%
of the average flux) was a substantial fraction of the Eddington
luminosity.
%%%%%%%%%%%%%%%%%%% FIGURE 1%%%%%%%%%%%%%%%%%%%%%%%%%%%%5
\begin{figure*}[htpb]
\epsscale{1} \plottwo{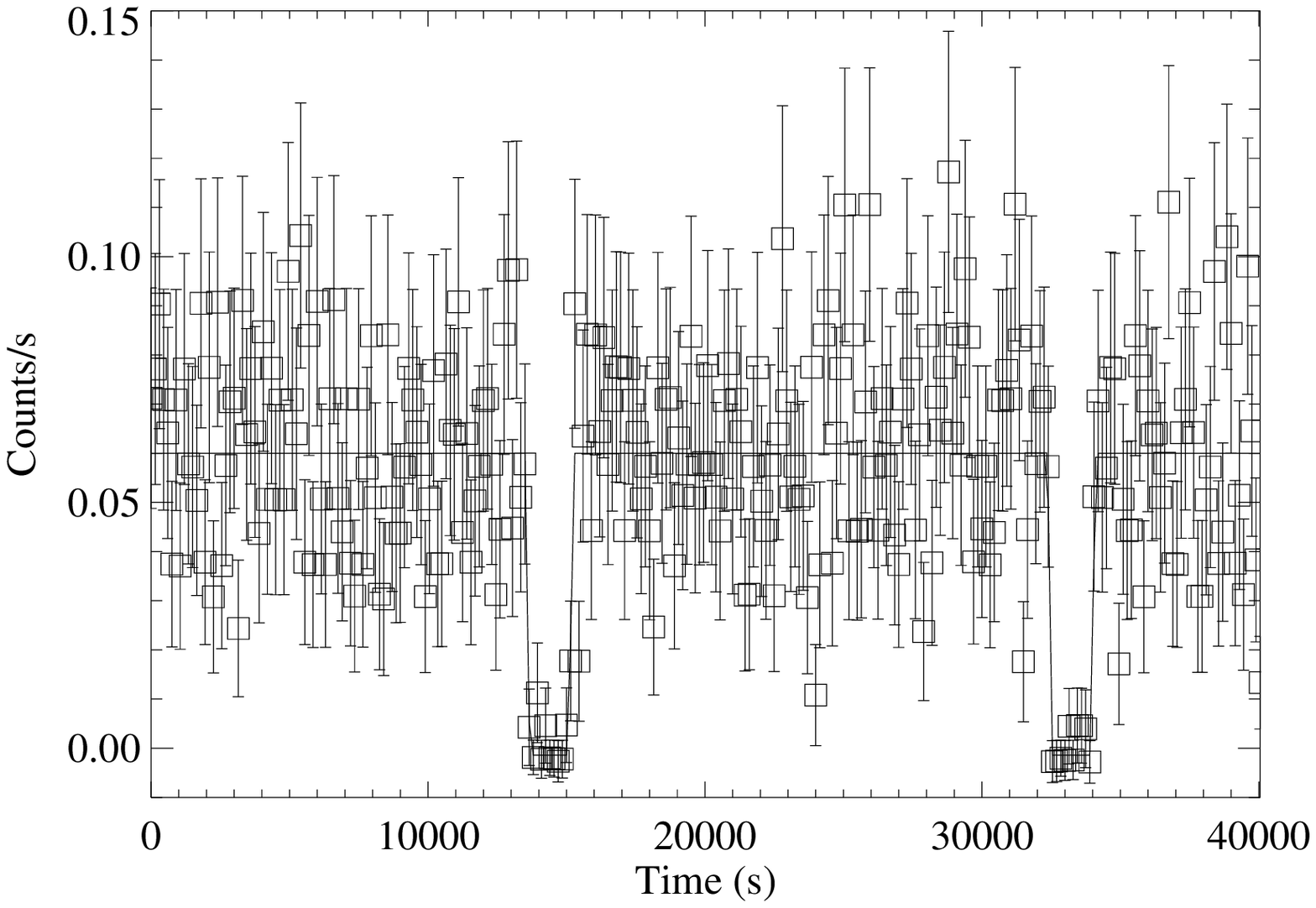}{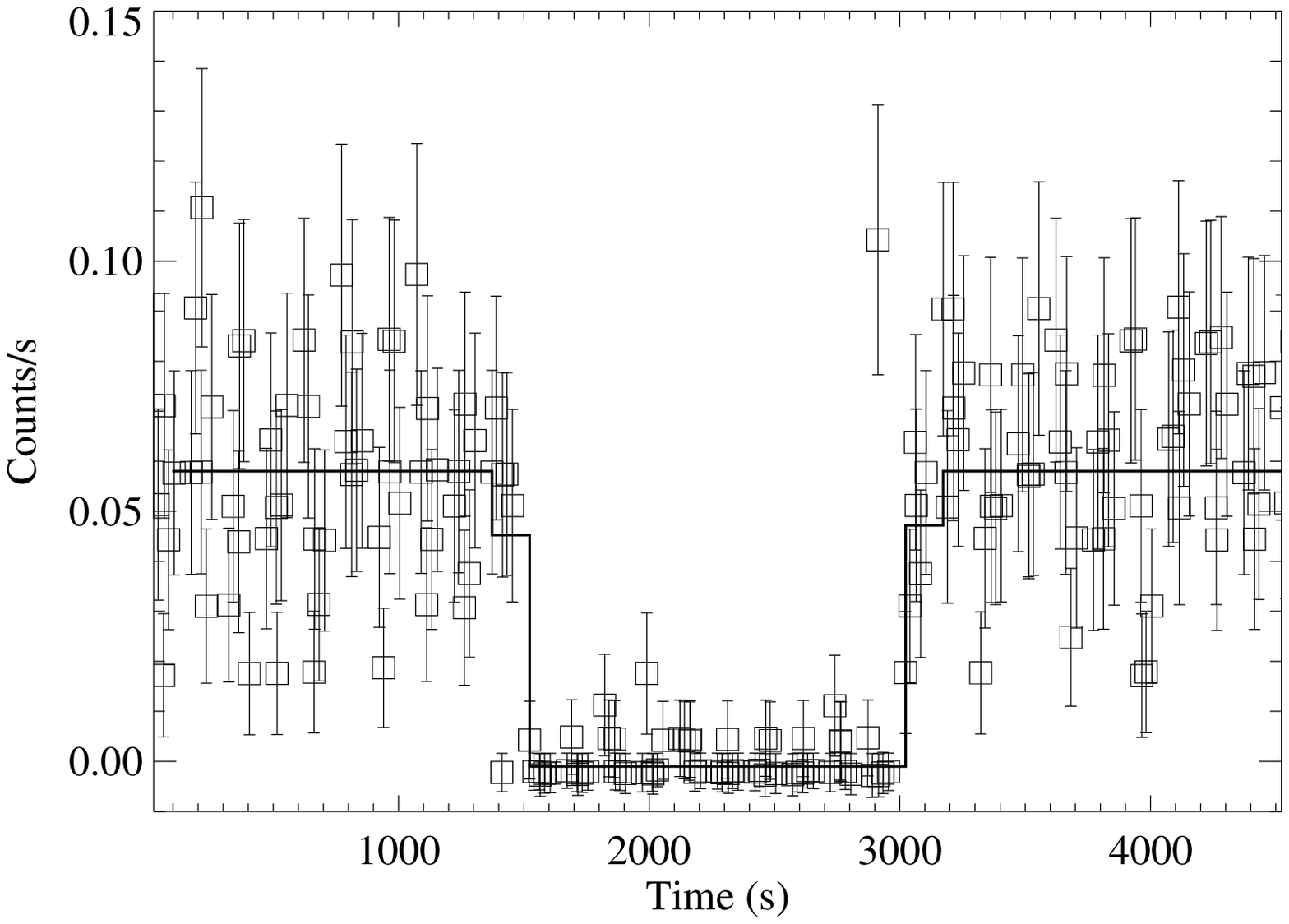}
 \caption {a) Left: Lightcurve from one of the 2007 {\it XMM-Newton}
   observations, fit by a total eclipse with finite duration ingress
   and egress.  No sinusoidal orbital modulation was evident. The
   $90\%$ confidence level upper limit of the modulation is $10\%$. b)
   Right: Lightcurve of six eclipses folded as one. The mid-eclipse
   epochs are aligned and centered at 2250\,s.
   \label{fig:fig1}}
\end{figure*}
%%%%%%%%%%%%%%%%%%%%%%%%%%%%%%%%%%%%%%%%%%%%%%%%%%%%%%%%%
The X-ray spectrum of 4U 2129+47, as determined with a {\it Chandra}
observation taken in 2000, was consistent with thermal emission plus a
powerlaw hard tail (N02). The 2--8\,keV flux was $\approx 40\%$ of the
0.5--2\,keV absorbed flux. The contribution of the ``powerlaw'' hard
tail to the 0.5--2\,keV band was, of course, model dependent, but was
consistent with being $\approx 20\%$ of the 0.5--2\,keV flux, as we
discuss below.  A sinusoidal orbital modulation (peak-to-peak
amplitude of $\pm30\%$) was observed in the X-ray lightcurve in the
same observation. The phase resolved spectra showed variation in the
hydrogen column density.  However, in the {\it XMM-Newton}
observations (B07), the sinusoidal modulation was absent, while the
flux of the powerlaw component was constrained to be less than $10\%$
of the $0.2-10$ keV flux (B07).

Here we report on recent {\it XMM-Newton} observations of
4U~2129+47. We combine analyses of these observations with
reanalyses of the previous observations, and show that, absent the
two 2005 {\it XMM-Newton} observations (B07), the historical
observations are consistent with a linear ephemeris with a
constant period. We outline our data reduction procedure in \S2
and present the timing and spectral analysis in \S3 and \S4
respectively. We summarize our conclusions in \S5.

\section{Observations and data}
{\it XMM Newton} observed 4U 2129+47 on Nov 29, Dec 20, 2007 and
on Jan 04, Jan 18, 2008. The total time span for each of these 4
observations was 43270\,sec, resulting in effective exposure times
of at least 30\,ksec for the EPIC-PN, EPIC-MOS1, and EPIC-MOS2
cameras for the first three observations. The remaining observing
time was discarded due to the background flares. The standard {\it
  XMM Newton} Science Analysis System (SAS
8.0) was used to process the observation data files (ODFs) and to
produce calibrated event lists. We used the EMPROC task for the
two EPIC-MOS cameras and used the EPPROC task for the EPIC-PN
camera. We used the high energy (E $>10$\,keV) lightcurves to
determine the Good Time Intervals, so as to obtain the event lists
that were not affected by background flares. The Good Time
Intervals are slightly different for EPIC-PN and EPIC-MOS cameras;
therefore, the overlap Good Time Interval was used to filter each
lightcurve with the EVSELECT keywords ``timemin" and ``timemax".

The low energy ($0.2-1.5$\,keV) source lightcurves and the spectra
($0.2-12$\,keV) were extracted within the circles of $14.6''$
radius centered on the source. Larger circles were not used in
order to avoid a Digital Sky Survey stellar object (labeled as
S$3-\beta$ in N02). We extracted background lightcurves and
spectra within the same CCD as the source region. The largest
source-free regions near the source, which were within circles of
radii about $116''$, were chosen for background extraction. The
SAS BACKSCALE task was performed to calculate the difference in
extraction areas between source and background. In order to obtain
the mid-eclipse epochs, we tried various bin sizes ($75$~s,
$150$~s and $300$~s etc.) to obtain the best compromise between
signal to noise ratio in a bin, and time resolution. Finally the
lightcurves were extracted with the bin size of $150$\,s. The
times of all lightcurves were corrected to the barycenter of the
Solar System by using the SAS BARYCEN task. Lightcurves from
EPIC-PN and two EPIC-MOS cameras were summed up with the LCMATH
task, in order to get better statistics in the fitting of the
eclipse parameters. Given the low count rate of the EPIC-MOS
cameras compared to the EPIC-PN camera, for the spectral analysis
we discuss only the spectrum from the EPIC-PN camera.

\section{Orbital ephemeris and eclipse parameters}

The observation made on Jan 18, 2008 was seriously affected by
background flares, however, 2 eclipses were found in each of the
other 3 observations respectively, resulting in 6 eclipses for
fitting.

The 3 pairs of eclipses were then simultaneously fit with the same set
of eclipse parameters. The ingress duration, the egress duration, and
the duration of the eclipse (defined to be from the beginning of the
ingress to the end of the egress) were set to be free parameters and
assumed to be the same for all 6 eclipses, under the hypothesis that
these eclipses should have the same shape (see \S5). The starting time
of the first eclipse ingress and the count rates within and out of the
eclipse were fit for each lightcurve individually. The separation
between the two connected eclipses within the same lightcurve was set
to be 18857.63\,s based upon the previous {\it Chandra} results.  The
second eclipse in each lightcurve therefore only contributed to the
fitting of the eclipse shapes, while only the first eclipse in each
lightcurve has been used to estimate the orbital ephemeris.  The
mid-eclipse epoch was defined to be the starting time of the ingress
plus half of the eclipse full ingress-to-egress duration.  The model
was integrated over each time bin, especially the bins that are
crossed by the ingress and the egress, in order to be compatible with
either the finite ingress/egress durations or the infinite
ingress/egress slopes (for a rectangular eclipse model).  $\chi^2$
minimization was performed in order to determine the best fit to the
lightcurve.

\newcommand\tabspace{\noalign{\vspace*{0.7mm}}}
\begin{deluxetable}{llll}
\setlength{\tabcolsep}{0.03in}
\tabletypesize{\footnotesize}
\tablewidth{0pt}
\tablecaption{Mid-eclipse epoch measurements.\label{tab:tab1}}
\tablehead{ \colhead{\textbf{Observatory}}
            & \colhead{Mid-eclipse Epoch}
            & \colhead{Orbital Period}
            & \colhead{References} \\
            & (JD)
            & (sec)
          }
\startdata
{\it Einstein}              &2444403.743(2)               &18857.48(7)              &(MC82)\\
%%%
\tabspace
%%%
{\it Chandra}               &2451879.5713(2)              &18857.631(5)             &(N02)\\
%%%
\tabspace
%%%
{\it XMM-Newton}            &2453506.4821(4)\tablenotemark{a}    &18857.638(7)      &(B07)\tablenotemark{a}\\
%%%
\tabspace
%%%
{\it XMM-Newton}            &2453528.3069(4)\tablenotemark{a}    &18857.636(7)      &(B07)\tablenotemark{a}\\
%%%
\tabspace
%%%
{\it XMM-Newton}            &2454433.8640(4)              &18857.632(7)             &this work\\
%%%
\tabspace
%%%
{\it XMM-Newton}            &2454455.6900(4)              &18857.632(7)             &this work\\
%%%
\tabspace
%%%
{\it XMM-Newton}            &2454470.0948(4)              &18857.631(7)             &this work\\
\enddata
\tablecomments{Numbers in parentheses are the errors (at 1$\sigma$
level) on the last significant digit.} \tablenotetext{a}{We
reprocessed the data for these two observations using a 150\,s
binned lightcurves in order to be consistent with our routine.}
\end{deluxetable}

%%%%%%%%%%%%%%%%%%% FIGURE 2%%%%%%%%%%%%%%%%%%%%%%%%%%%%5
\begin{figure*}[htbp]
\begin{center}
\epsscale{1.15} \plottwo{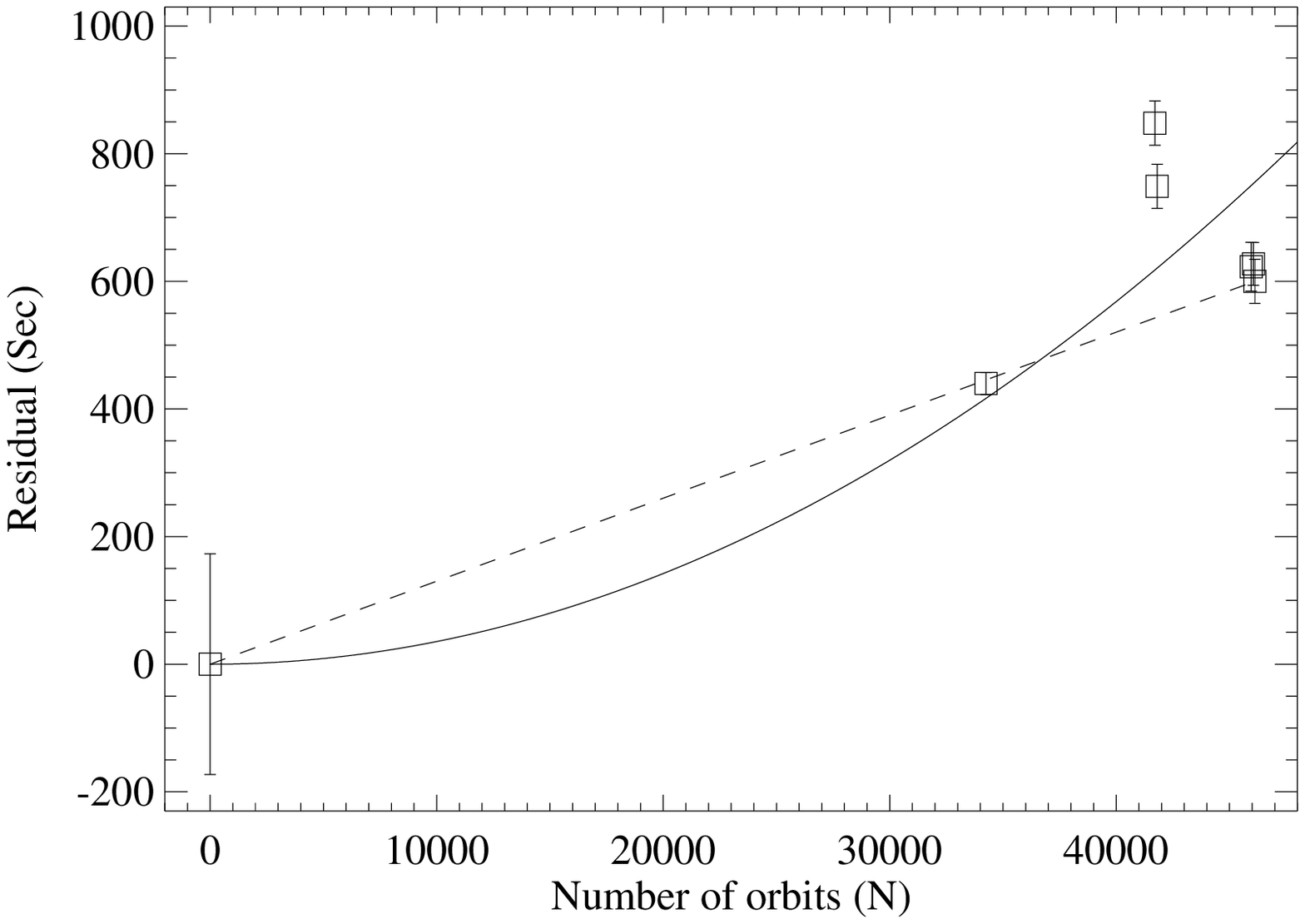}{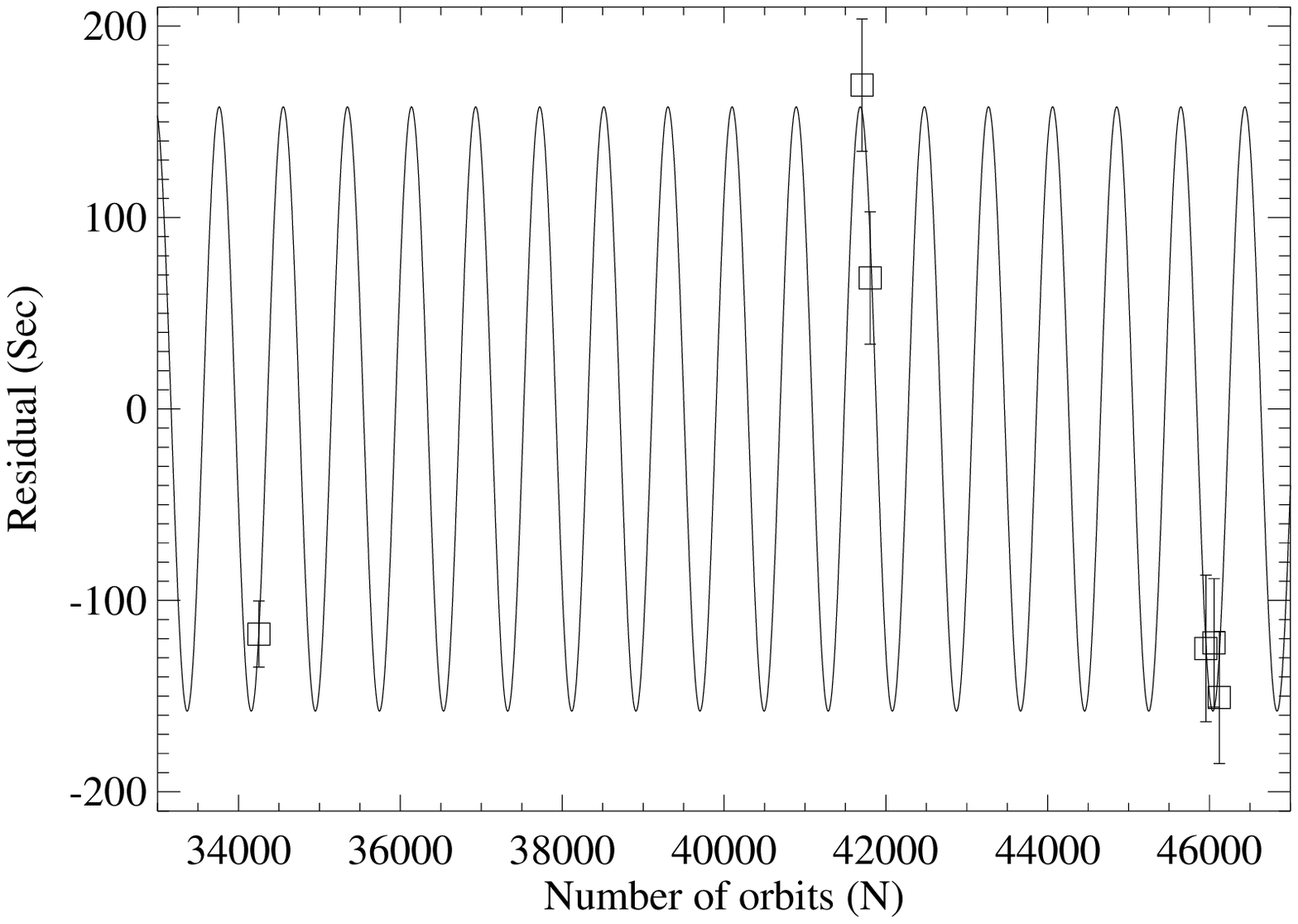}
 \caption {\label{fig:fig2} Each square shows the timing
   residual (defined as the difference between the observed ephemeris
   and the predicted ephemeris) at certain number of orbits from the
   observations listed in Table.~\ref{tab:tab1}. a) Left:  The dashed line denotes
   a linear ephemeris with a period of $18857.63$, while the solid line
   denotes a quadratic ephemeris. Absent the 2005 {\it XMM-Newton} observations,
   the ephemeris of the eclipses can be well fitted by
   a constant period. However, we confirmed the strong deviations of
   the 2005 {\it XMM-Newton} observations from any linear or quadratic
   ephemeris. b) Right: Close up of the third body fit with a third body
   period of $\sim 175$~days, showing just the {\it Chandra} and
   {\it XMM-Newton} data since year 2000. The solid line denotes a sinusoidal ephemeris. }
\end{center}
\end{figure*}
%%%%%%%%%%%%%%%%%%%%%%%%%%%%%%%%%%%%%%%%%%%%%%%%%%%%%%%%%
One of our fitted lightcurves is shown in Fig.~\ref{fig:fig1}a.  The
best fits to the mid-eclipse epochs were found to be
$T_{0}(1)=2454~433.8640\pm 0.0004$~JD, $T_{0}(2)=2454~455.6900\pm
0.0004$~JD and $T_{0}(3)=2454~470.0948\pm 0.0004$~JD, with
$\chi^2/$d.o.f.=$782.6/708$ (errors are at $68\%$ confidence
level). The best fit result favors a rectangular eclipse model with
very short duration of the ingress and egress durations; upper limits
of 50\,s and 30\,s were found for the ingress and egress durations
respectively. A folded lightcurve of 6 eclipses is shown in
Fig.~\ref{fig:fig1}b, where the mid-eclipse epochs are aligned and
centered. The sinusoidal variation seen in the {\it Chandra}
observation is absent; the upper limit of the modulation amplitude is
$< 10\%$ at $90\%$ confidence level. The duration of the eclipses were
derived to be $1565\pm23$\,s. We also re-analyzed the previous two
{\it XMM-Newton} observations of May 15 and June 6, 2005 with the
above methods in order to get a direct comparison with our data
sets. These two lightcurves were also extracted with 150\,s bins, and
there was one eclipse present in each of them. The best fit
mid-eclipse epochs are $T_{0}(a)=2453~506.4821\pm 0.0004$~JD and
$T_{0}(b)=2453~528.3069\pm 0.0004$ with $\chi^2/$d.o.f.=$18.8/21$ and
$87.8/84$, respectively.

We considered the above mid-eclipse epochs together with epochs,
$T_{n}$, derived from the previous observations
(Table~\ref{tab:tab1}) in order to determine a refined orbital
solution and to measure any orbital period derivative (which was
weakly suggested by the analysis of N02). We considered the
ephemeris from MC82 as reference ($T_{ref}=2444~403.743\pm
0.002$~JD, $P_{ref}=18~857.48\pm 0.07$\,s), and calculated $n$,
the closest integer to $(T_{n}-T_{ref})/P_{ref}$. Our three
observations (the first eclipse in each of the three lightcurves)
correspond to $n= 45~955, 46~055, 46~121$. The average orbital
periods ($P=(T_{n}-T_{ref})/n$) inferred from these observations
are therefore $18857.632\pm 0.005$\,s, $18857.631\pm 0.005$\,s and
$18857.632\pm 0.004$\,s. We noticed that, absent the 2005 {\it
XMM-Newton} observations, the 2007/2008 observations are
consistent with a linear ephemeris with a constant period of
$18~857.63$\,s.  However, we confirmed the strong deviations of
the 2005 {\it XMM-Newton} observations from any linear or
quadratic ephemeris (Fig.~\ref{fig:fig2}). We therefore tried to
fit the linear timing residual by a third body orbit on top of a
steady binary orbital period. For simplicity, we assume the
interaction between a binary system and a third body results in a
sinusoidal residual.  The ephemeris is not uniquely determined; we
found valid ephemerides for the third body orbit at $\sim 800$,
1300, 1700, 1900, and 2900 times the binary orbit. The shortest
consistent sinusoidal period was therefore found to be $\sim
175$\,days (Fig.~\ref{fig:fig2}).

\section{Spectral Analysis}
%%%%%%%%%%%%%%%%%%% FIGURE 3%%%%%%%%%%%%%%%%%%%%%%%%%%%%5
\begin{figure*}[htpb]
\epsscale{1.1} \plotone{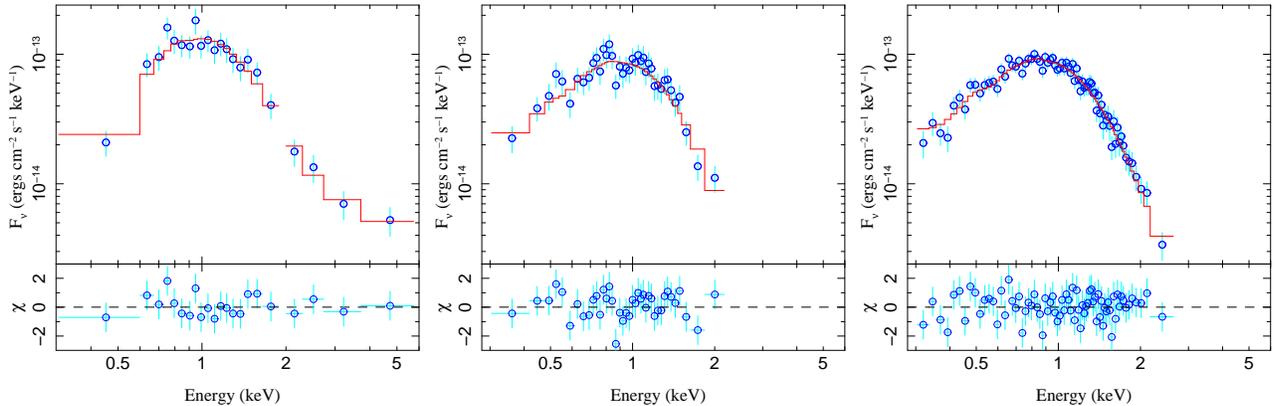}
 \caption {Combined spectra for the {\it Chandra} observations of
   Nowak, Heinz, \& Begelman (2002) (left), the {\it XMM-Newton}
   observations of Bozzo et al. (2007) (middle), and the three {\it
     XMM-Newton} observations presented in this work (right).  Each
   has been fit with a model consisting of an absorbed blackbody plus
   powerlaw; however, the photon index is fixed to $\Gamma=2$ for the
   {\it XMM-Newton} observations.  Note that all spectra \emph{have
     been unfolded without reference to the underlying fitted spectral
     model}.  (The unfolded spectra do, however, reference the
   response matrices of the detectors; see \protect{\citealt{now05}}
   and the Appendix for further details.) \label{fig:fig3}}
\end{figure*}
%%%%%%%%%%%%%%%%%%%%%%%%%%%%%%%%%%%%%%%%%%%%%%%%%%%%%%%%%
The spectra of the 3 observations were accumulated during the same
time intervals selected for the extraction of the EPIC-PN
lightcurves, except that the eclipses were also excluded. Spectral
analysis was carried out using ISIS version 1.4.9-55
\citep{houck00}.  Since there is no evidence of orbital modulation
in the lightcurves, the spectra we produced were not phase
resolved. Furthermore, preliminary fits indicated little or no
variability among spectra from individual observations; therefore,
we fit all spectra simultaneously. Specifically, we used the ISIS
{\tt combine\_datasets} functionality \footnote{This is
essentially equivalent to summing the
  pulse height analysis (PHA) files and summing the product of the
  response matrix and effective area files.  Whereas this can be
  accomplished, e.g., using {\tt FTOOLS} outside of the analysis
  program (and in fact, ``outside of analysis'' is the only mode
  supported by {\tt XSPEC}), performing summing during analysis allows
  one to examine data and residuals for each data set individually,
  choose different binning and noticing criteria based upon individual
  spectra, etc.} to sum the spectra and responses during analysis.
The data were binned to have a minimum signal-to-noise
\footnote{To be
  explicit, here and throughout we mean that the source counts,
  excluding the estimated background counts, divided by the estimated
  error, including that of the background counts, meets or exceeds the
  given signal-to-noise threshold in each spectral bin.} of 4.5 and a
minimum of four energy channels per grouped spectral bin. The
spectra were fit over an energy range such that the lower bounds
of the spectral bins exceeded 0.3\,keV, and the upper bounds of
the spectral bins were less than 8\,keV.  (This latter criteria,
with the above grouping, effectively restricted the upper bound of
the spectra to be $\approx 3$\,keV.)

The spectra could be well-fitted by an absorbed blackbody, both with
and without a powerlaw (Table~\ref{tab:tab2}).  For the absorption,
following the fits of N02, we used the model of
\citet{wilms:00}. Given that the presence of any powerlaw was not well
constrained, we froze its photon index to $\Gamma=2$ when including
such a component in the fits.  Without the additional powerlaw, the
best fit temperature and column density were
$0.21^{+0.01}_{-0.01}$\,keV and
$(0.21\pm{0.02})\times10^{22}$\,cm$^{-2}$, with a $\chi^2=62.6/$73
degrees of freedom. With the additional powerlaw, the best fit
temperature and column density were $0.20^{+0.01}_{-0.01}$\,keV and
$(0.24\pm0.03)\times10^{22}$\,cm$^{-2}$, with a $\chi^2=57.8/$73
degrees of freedom.  Both of these sets of parameters are consistent
with the thermal component of the fits presented by N02 for the
\emph{peak} (i.e., least absorbed part) of the sinusoidal modulation
of the {\it Chandra} lightcurve.  We show the spectra and fit
including the powerlaw in Fig.~\ref{fig:fig3}.

As indicated by these fits (all of which have reduced $\chi^2<1$), the
powerlaw hard tail is no longer required to describe the spectrum.
Including a powerlaw slightly increases the best fit ${\rm N_H}$ and
its associated error bars, while slightly decreasing the best fit
temperature.  Our best fit models, with or without a powerlaw, yield
an absorbed 0.5--2\,keV flux of $(7.7\pm0.1)\times10^{-14}\,{\rm
  ergs~cm^{-2}~s^{-1}}$, with no more than 10\%, i.e.,
$0.7\times10^{-14}\,{\rm ergs~cm^{-2}~s^{-1}}$ (90\% confidence level)
being attributable to a $\Gamma=2$ powerlaw.  This 0.5--2\,keV flux is
36\% lower than the $(1.2\pm0.1)\times 10^{-13}\,{\rm
  ergs~cm^{-2}~s^{-1}}$ found during the peaks of the {\it Chandra}
lightcurve (see below).  Note that above, and throughout this work,
unless stated otherwise we will quote 68\% confidence limits for
fluxes, but 90\% confidence limits for fit parameters.

To estimate the 2--6\,keV flux, we grouped the spectrum (starting at
2\,keV) to have a minimum signal-to-noise of 3 and a minimum of 2
channels per grouped energy bin, and then we fit an unabsorbed
powerlaw spectrum between 2--6\,keV.  (Grouping to higher
signal-to-noise left no channels between $\approx 3.5$--6\,keV.) This
yields an estimated $(6\pm3)\times10^{-15}\,{\rm ergs~cm^{-2}~s^{-1}}$
in the 2--6\,keV band, which is to be compared to the nearly five
times larger 2--6\,keV flux of $(2.9\pm0.6)\times10^{-14}\,{\rm
  ergs~cm^{-2}~s^{-1}}$ from the {\it Chandra} observations (see
below).

The question then arises as to when the drop in the 0.5--2\,keV flux
occurred, and whether or not it is solely attributable to the loss of
the hard X-ray tail. To further explore these issues, we applied
absorbed blackbody plus powerlaw fits to the spectra described by B07
(out of eclipse data only, grouped and noticed exactly as for the
2007/2008 {\it XMM-Newton} spectra described above, with powerlaw
index frozen to $\Gamma=2$), and refit the spectra of N02 (powerlaw
index left unfrozen). We used the same exact data extractions and
spectral files from N02. For the {\it Chandra} observations, we follow
N02 and fit the 0.3--2\,keV spectra (grouped to signal-to-noise of 4.5
at 0.3\,keV and above) from the peak of the lightcurve's sinusoidal
modulation, while we fit the 2--6\,keV spectra (grouped to a
signal-to-noise of 4 at 2\,keV and above) from all of the out of
eclipse times.  The flux levels quoted above correspond to these new
fits of the {\it Chandra} data.

The fits to the data from B07 are completely consistent with the
fits to the new data, and yield a blackbody temperature and
neutral column of $0.21^{+0.01}_{-0.03}$\,keV and
$0.27^{+0.09}_{-0.04} \times 10^{22}\,{\rm cm^{-2}}$,
respectively, with $\chi^2=32.2/37$ degrees of freedom.  The
powerlaw normalization was consistent with 0, and thus excluded a
powerlaw from the fits to these data.  The absorbed 0.5--2\,keV
flux was $(7.5\pm0.3)\times10^{-14}\,{\rm
  ergs~cm^{-2}~s^{-1}}$, i.e., comparable to that from the more recent
{\it XMM-Newton} data.  The spectra are shown in
Fig.~{\ref{fig:fig3}}.

As discussed in N02, and shown in Fig.~\ref{fig:fig3}, the {\it
  Chandra} data clearly indicate the presence of a hard tail.  Here we
find ${\rm N_H} = 0.39^{+0.18}_{-0.11} \times 10^{22}\,{\rm cm^{-2}}$,
$kT = 0.21^{+0.03}_{-0.04}$\,keV, and $\Gamma=2.0^{+1.5}_{-1.6}$, with
$\chi^2=10.7/17$ degrees of freedom.  Note that the larger ${\rm N_H}$
value found here compared to N02 is due to the inclusion of the
0.3--0.5\,keV data, and is partly indicative of a systematic
dependence of this parameter upon the index of the fitted powerlaw.
For the best-fit powerlaw index, $0.2 \times10^{-14}\,{\rm
  ergs~cm^{-2}~s^{-1}}$ of the 0.5--2\,keV absorbed flux is
attributable to the powerlaw.  This is not enough to account for the
drop in flux between the {\it Chandra} and {\it XMM-Newton}
observations.  However, if for the {\it Chandra} spectra we fix the
powerlaw at its 90\% confidence level upper limit ($\Gamma = 3.5$),
then $0.6 \times10^{-14}\,{\rm ergs~cm^{-2}~s^{-1}}$ of the
0.5--2\,keV absorbed flux is attributable to the powerlaw component
(albeit with an ${\rm N_h} = 0.58 \times 10^{22}\,{\rm cm^{-2}}$,
which is higher than the best fit values for the {\it XMM-Newton}
data).  This is more than enough to account for all of the change in
the 0.5--2\,keV absorbed flux, and highlights some of the systematic
uncertainties inherent in determining bolometric flux changes (i.e.,
the need to adequately model the changes in the local column and to
use realistic models for the hard X-ray tail).  With these current
data, it is difficult to distinguish between resumed cooling of the
neutron star thermal component, or mere loss of the additional
(presumed external) hard tail component that here is modeled with a
powerlaw.

%%%%%%%%%%%%%%%%%%% FIGURE 4%%%%%%%%%%%%%%%%%%%%%%%%%%%%
\begin{figure}[htpb]
\epsscale{1} \plotone{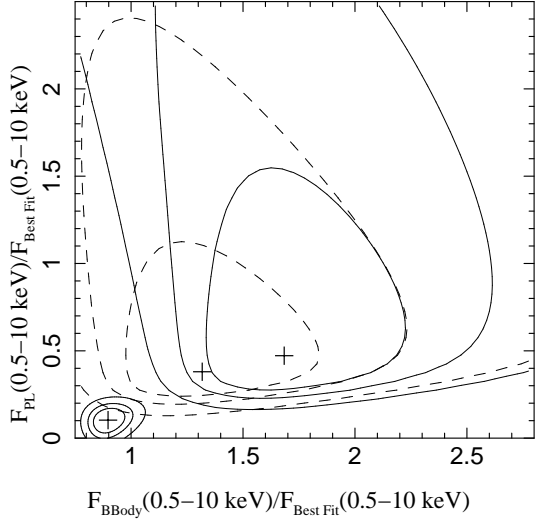}
 \caption {68\%, 90\%, and 99\% confidence contours for the
   \emph{unabsorbed} 0.5--10\,keV flux in the blackbody ($x$-axis) and
   powerlaw ($y$-axis) components from fits to the 0.3--6\,keV spectra
   from the 2007/2008 {\it XMM-Newton} observations (lower left) and
   {\it Chandra} observation (upper right). The dashed confidence
   contours correspond to fits to the 0.6--6\,keV band of the {\it
     Chandra} spectra.  Plus signs indicate the ``best fit'' values.
   Note that all flux values are shown relative to the total,
   unabsorbed flux in the 0.5--10\,keV band determined from the best
   blackbody plus powerlaw fit to the 2007/2008 {\it XMM-Newton}
   observations. \label{fig:fig4}}
\end{figure}
%%%%%%%%%%%%%%%%%%%%%%%%%%%%%%%%%%%%%%%%%%%%%%%%%%%%%%%%%

To highlight some of the issues with determining bolometric flux
changes and with attributing any changes to specific model components,
in Fig.~\ref{fig:fig4} we show error contours for the
\emph{unabsorbed} 0.5--10\,keV flux in the blackbody and powerlaw
components when fitting the 2007/2008 {\it XMM-Newton} spectra and the
{\it Chandra} spectra.  These errors account for uncertainties in the
neutral column, but do not address systematic issues with the choice
of model itself.  We see that formally the 99\% confidence level
contours \emph{do not overlap}, which would indicate that \emph{both}
the thermal and powerlaw components have decreased between the time of
the {\it Chandra} and {\it XMM-Newton} observations.  However, if we
ignore the first bin (0.3--0.6\,keV) in the {\it Chandra} spectra, the
{\it Chandra} contours shift significantly leftward towards lower
blackbody flux.  (Additionally, the fitted neutral column also
decreases.)  There is less of a shift downward in powerlaw flux
between these two fits of the {\it Chandra data}.  Thus, as stated
above, it remains somewhat ambiguous as to what extent the changes
between the {\it Chandra} and {\it XMM-Newton} data can be attributed
solely to components other than the thermal emission from the neutron
star.

%%%%%
%%%%%  CHANGES BEGIN HERE
%%%%%

%%%%%%%%%%%%%%%%%%% FIGURE 5%%%%%%%%%%%%%%%%%%%%%%%%%%%%
\begin{figure}[htpb]
\epsscale{1} \plotone{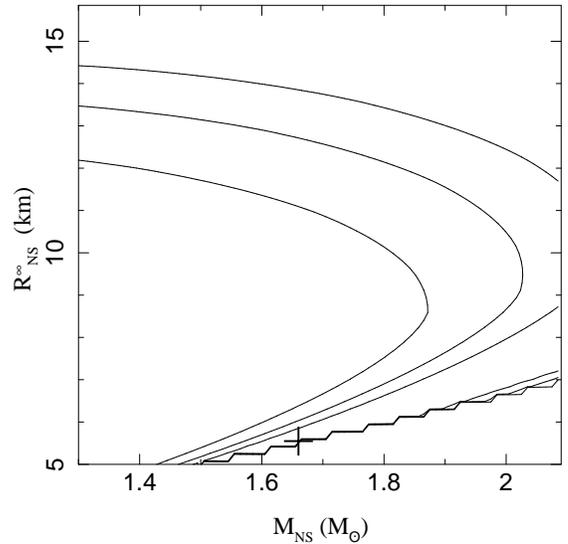}
 \caption {68\%, 90\%, and 99\% confidence contours for the neutron
   star mass ($x$-axis) and radius ($y$-axis) from neutron star
   atmosphere ({\tt NSATMOS}) model fits to the 0.3--6\,keV spectra from
   the 2007/2008 {\it XMM-Newton} observations.  The neutron star
   distance was fixed to 6.3\,kpc. The plus sign corresponds to the
   best fit. The jagged contours at the bottom of the figure represent the
   limits of the interpolation grids used in the calculations of the
   NSATMOS model.
   \label{fig:fig5}}
\end{figure}
%%%%%%%%%%%%%%%%%%%%%%%%%%%%%%%%%%%%%%%%%%%%%%%%%%%%%%%%%

We have fit also the {\it XMM-Newton} spectra with the neutron
star atmosphere ({\tt NSA}) model of \citet{zavlin:96a}, by fixing
the neutron star distance to $6.3$\,kpc and the neutron star mass
to 1.4\,$M_\odot$.  Results of these fits are presented in
Table~\ref{tab:tab2}.  They are consistent with those of N02 and
B07 (although N02 also required the inclusion of a powerlaw
component). Specifically, we find neutron star radii of $\approx
5$\,km, and effective temperatures in the range of
2--3$\times10^6$\,K. If we instead use the {\tt NSATMOS} model of
\citet{heinke:06a}, again fixing the neutron star mass and
distance as above, we find lower neutron star temperatures
($\approx 10^6$\,K) and significantly larger radii ($> 12$\,km).
This raises the question of the degree to which one can find a
consistent fitted mass and radius among all the datasets discussed
here.

To explore this question, we have performed a joint fit of the
{\it Chandra} and {\it XMM-Newton} spectra. The observations from
the individual observing epochs (2000, 2005, 2007/2008) were
grouped and added as described above, with the additional caveat
that we now exclude the {\it Chandra} data below 0.6\,keV so as to
minimize the influence of the soft-end of the powerlaw component
on the fitted neutral column. We include a hard tail, modeled as a
$\Gamma=2$ powerlaw, in fits to the {\it Chandra} spectra, and
again use the {\tt
  NSATMOS} model \citep{heinke:06a} to describe the soft spectra from
all epochs.  For this latter component we again fix the distance to
6.3\,kpc and we further constrain the neutron star mass and radius to
be the same for all epochs.  The individual epochs, however, are
fitted with independent neutron star temperatures and neutral columns.
Results are presented in Table~\ref{tab:tab3}.

We find that one can find a set of consistent parameters that are
statistically acceptable. There is a modest need for the {\it
  Chandra} spectra to be described with a slightly larger neutral
column than that fit to the 2007/2008 {\it XMM-Newton} spectra.  The
{\it Chandra} spectra also require a slightly higher neutron star
temperature, but here all temperatures fall within each others error
bars.  The best fit mass is 1.66\,${\rm M_\odot}$ and the best fit
radius is 5.55\,km.  This is somewhat smaller than the values
presented in Table~\ref{tab:tab2} when applying the {\tt NSATMOS}
model with a fixed mass of 1.4\,${\rm M_\odot}$, and is more
consistent with the values found when using the {\tt NSA} model.

The reasons for this can be elucidated by examining the error
contours for the fitted mass and radius. In Fig.~\ref{fig:fig5} we
show the mass/radius contours obtained from applying an absorbed
{\tt NSATMOS} model to just the 2007/2008 {\it XMM-Newton} data
(i.e., our best measured spectrum, with no discernible hard tail).
We see that the error contours admit a wide range of masses and
radii, with the curvature of these contours generally favoring two
regimes: small radius (with high temperature - not shown in this
figure) and large radius (with lower temperture). Formally, the
small radius/high temperature solution is the statistical minimum;
however, it is not statistically very different than the large
radius/low temperature regime.  (Comparable contours for the {\tt
NSA} model look very similar, albeit with the contours shifted to
lower mass values.)

These different fit parameter regimes add further systematic
uncertainty to estimates of the unabsorbed, 0.5-10\,keV flux. As
shown in Tables~\ref{tab:tab2} and \ref{tab:tab3}, the estimate of
this flux ranges from 1.5--1.9$\times10^{-13}\,{\rm
erg~cm^{-2}~s^{-1}}$ for the 2007/2008 {\it XMM-Newton} data
alone. Furthermore, we see that depending upon whether we use a
phenomenological model (the blackbody plus powerlaw fits)
individually fit to each epoch, or a more physical model (e.g.,
{\tt NSATMOS}) jointly fit to all epochs, we either find evidence
for further neutron star cooling or evidence of a consistent
neutron star surface temperature. We again caution, however, that
determining the bolometric flux in specific model components can
be fraught with many systematic uncertainties (e.g.,
Fig.~\ref{fig:fig4}), and we therefore consider any evidence for
further neutron star cooling to be ambiguous at best.

%%%%%
%%%%%  CHANGES END HERE
%%%%%

\section{Conclusions}

We reported on recent {\it XMM-Newton} observations of 4U 2129+47 in a
quiescent state. Given lower background flaring rates and relatively
longer durations of these observations compared to the 2005 {\it
  XMM-Newton} observations reported by \citet{boz07}, we were able to
make clear the following three points:

\begin{enumerate}
\item The 0.5--2\,keV absorbed X-ray flux of the source has been
  reduced by more than $\approx 40\%$ compared to the {\it Chandra}
  observation. This was even true for the observations discussed by
  B07.

\item The sinusoidal variation seen in the {\it Chandra} observation
  ($\pm30\%$ peak-to-peak) is absent, or at least greatly diminished
  ($90\%$ confidence level upper limits of $\pm10\%$ modulation).
  Additionally, the associated neutral column exhibits no orbital
  variability, and is consistent with the brightest/least absorbed
  orbital phases of the {\it Chandra} observation.

\item The powerlaw tail seen in the {\it Chandra} observation is
  absent. A $\Gamma=2$ powerlaw contributes less than $10\%$ of the
  0.5--2\,keV flux, and is not evident in the $2-6$~keV band. The
  2--6\,keV flux is reduced by a factor of five compared to the {\it
    Chandra} observation.
\end{enumerate}

With this further drop in the X-ray flux of the source, we may now
truly be seeing 4U~2129+47 enter into a quiescent stage solely
dominated by neutron star cooling, with little or no contribution
from residual weak accretion. The spectrum of the source can be
well explained by absorbed blackbody emission from the surface of
a neutron star, with an emission radius of $\sim1.8$~km. Likewise,
more sophisticated neutron star atmosphere models also describe
the spectra very well, with no need for an additional component.

As discussed by N02, the sinusoidal variability detected with the
{\it Chandra} observation could have been due to a neutral
hydrogen column being raised by the interaction of the accretion
stream with the outer edge of an accretion disk
\citep{par00,hei01}. The fact that the sinusoidal modulation was
absent in the 2007, and likely also the 2005 {\it XMM-Newton}
observations could suggest that the system is in a lower accretion
state, and might indicate that the geometry of the outer disk
region has changed. It has been hypothesized that the hard tails
of quiescent neutron stars originate from a pulsar wind, only able
to turn on in quiescence, that is interacting with the accretion
stream at large radius \citep{cam98,cam00}. The vanishing of both
the hard tail and the sinusoidal modulation in 4U~2129+47 could
suggest that what we had observed in the {\it Chandra} observation
was indeed the interaction between the pulsar wind and the disk
edge, perhaps as the last remnants of a disk accreted onto the
neutron star. This is consistent with the fact that for these
current {\it XMM-Newton} observations the fitted neutral column is
$\approx 2$--$3\times10^{21}\,{\rm cm}^{-2}$, which is comparable
to the value measured for the least absorbed/peak of the {\it
Chandra} lightcurve. We note that a dominant nonthermal component
has been reported in a ``recycled'' system: the binary radio
millisecond pulsar PSR J0024$-$7204W. The X-ray variability
observed in PSR J0024$-$7204W suggests that the hard X-ray
emission is produced by interaction between the pulsar wind and
matter from the secondary star, which occurs much closer to the
companion star than the millisecond pulsar \citep{bog05}.  The
previously observed hard tail in 4U~2129+47 might have been a
similar phenomenon.

What remains ambiguous is whether the associated drop in the
0.5--2\,keV absorbed flux is solely related to the vanishing of the
hard tail, or whether additional cooling of the neutron star has
occurred, or continues to occur.  Although the soft X-ray flux has
clearly declined from the {\it Chandra} observations to the {\it
  XMM-Newton} observations, there is no evidence for a decline between
the 2005 and 2007/2008 observations.  The initial drop of the soft
X-ray flux is consistent with solely the loss of the hard tail for
some parameter regimes of this component; however, for the best-fit
parameters, additional cooling would have had to occur.

We note that the expected luminosity of 4U~2129+47 due to neutron star
cooling has recently been discussed by \citet{heinke:08a}.  Even
though 4U~2129+47 is among the brighter of the 23 quiescent sources
presented by \citet{heinke:08a}, the degree to which it is ``too
faint'', and thus requires ``non-standard'' cooling (kaon cooling,
pion cooling, etc.; see \citealt{heinke:08a} and references therein)
relies on the rather uncertain average heating history of 4U~2129+47,
and the knowledge that the bulk of the observed soft X-ray emission
was \emph{not} due to residual accretion.  We note that any implied
reduction of the soft X-ray emission of 4U~2129+47, or any continued
cooling observed in the future, increases the need to invoke
``non-standard'' cooling mechanisms for this system, and further
limits the parameter space for the long term average heating that
would allow for ``standard'' cooling.

A question also arises as to whether or not the loss of the hard X-ray
tail is in any way related to the unusual timing residuals associated
with the 2005 {\it XMM-Newton} observations (B07).  Strong deviations
from a simple linear or quadratic ephemeris have also been noted for
the bursting neutron star system EXO~0748$-$676 \citep{wolff:09a},
with these changes possibly being correlated with the duration of the
eclipse as viewed by the {\it Rossi X-ray Timing Explorer}
(RXTE). (Note that comparable eclipse duration changes, $<20$\,s,
observed in EXO~0748$-$676 are too small to have been seen in
4U~2129+47 with our observations.) \citet{wolff:09a} hypothesize that
these eclipse duration and ephemeris changes are related to magnetic
activity within the secondary.  We note, however, that over a
comparable five year span, the peak-to-peak residuals of
EXO~0748$-$676 (70\,s) are three to four times smaller than the
peak-to-peak residuals seen in 4U~2129+47 (200--300\,s).  Furthermore,
the trend of ephemeris timing residuals of EXO~0748$-$676 are more
persistently in one direction (although deviations do occur) than
those that we show in Fig.~\ref{fig:fig2} (compare to Fig.~4 of
\citealt{wolff:09a}).

In our work, we confirmed the strong deviations of two 2005
{XMM-Newton} eclipses from any linear or quadratic ephemeris, however,
this deviation has only been observed once since 1982. Rather than
presuming ephemeris residuals comparable to those of EXO~0748$-$676,
given the optical observations \citep[e.g.][]{gar87,both08}, we
instead have adopted the hypothesis of a triple system.  Given the
lack of constraints, however, we had to assume a sinusoidal function
for simplicity. With this assumption, the chance of detecting a
maximal time deviation and a minimal deviation should be equal. The
rare presence of the strong deviation may be due to a large
eccentricity of the third body orbit, so that the third star rarely
interacts strongly with the inner binary system. The third body
orbital period was estimated to be around 175 days, but can not be
uniquely determined with the current observations. In order to confirm
and accurately determine the orbital period of the third body, further
observations are required. Such future observations could also address
whether the soft X-ray flux of 4U 2129+47 has continued to drop, and
whether the hard tail and sinusoidal modulation remain absent.

\clearpage

\begin{landscape}

\begin{deluxetable}{lcccccccc}
\setlength{\tabcolsep}{0.01in}
\tabletypesize{\tiny} \tablewidth{0pt} \tablecaption{Best-fit
spectral parameters \label{tab:tab2}}
\tablehead{\colhead{\textbf{Parameters}}   &
\multicolumn{4}{c}{XMM-Newton (2007/2008)}
&\multicolumn{3}{c}{XMM-Newton (2005)}
            & Chandra \\
            & (no powerlaw)\tablenotemark{a} & (no powerlaw)\tablenotemark{b}  & ($\Gamma=2$, fixed) & (no powerlaw) & (no powerlaw)
            & (no powerlaw)\tablenotemark{a} & (no powerlaw)\tablenotemark{b} & ($\Gamma$ free)
          }
\startdata ${\rm N_{H}}$ ($10^{22}$~cm$^{-2}$)
&$0.21^{+0.02}_{-0.03}$ &$0.24\pm0.03$ & $0.29\pm0.03$ &
$0.32\pm0.03$
& $0.26^{+0.07}_{-0.06}$  & $0.33^{+0.11}_{-0.06}$   & $0.37\pm0.07$ & $0.39^{+0.18}_{-0.11}$   \\
%%%
\tabspace
%%%
$kT_{bb}$ (keV)
&$0.213^{+0.007}_{-0.006}$ &$0.202^{+0.011}_{-0.010}$  & \nodata &
\nodata
& $0.210^{+0.015}_{-0.015}$  & \nodata  & \nodata & $0.210^{+0.033}_{-0.043}$   \\
%%%
\tabspace
%%%
$R_{bb}$ (km)\tablenotemark{c}            &$1.8\pm0.2$
&$2.0^{+0.4}_{-0.2}$   & \nodata  & \nodata
& $1.9^{+0.6}_{-0.4}$  & \nodata  & \nodata & $2.5^{+2.3}_{-0.9}$   \\
%%%
\tabspace
%%%
A$_{\rm PL}$ ($10^{-5}~{\rm
\gamma/keV/cm^{2}/sec}$)\tablenotemark{d}
                                         & \nodata                &$0.3^{+0.3}_{-0.2}$    & \nodata & \nodata
& \nodata  & \nodata  & \nodata & $1.5^{+7.4}_{-1.3}$   \\

%%%
\tabspace
%%%
$\Gamma$ & \nodata & {\it 2} & \nodata & \nodata & \nodata & \nodata & $2^{+1.5}_{-1.6}$ \\
%%%
\tabspace
%%%
$\log T_{eff}$ (log K)  & \nodata & \nodata & $6.37\pm0.06$ &
$6.00\pm0.03$
& \nodata  & $6.39^{+0.16}_{-0.13}$   & $5.99\pm0.06$ &  \nodata  \\
%%%
\tabspace
%%%
$R^\infty_{NS}$ (km)\tablenotemark{e}  & \nodata & \nodata &
$5.3^{+0.1}_{-0.3}$ & $16.7^{+2.8}_{-2.6}$
& \nodata  & $5.2^{+1.5}_{-0.2}$  & $18.2^{+7.6}_{-5.8}$ &  \nodata \\
%%%
\tabspace
%%%
$\chi^2$/d.o.f.            &62.6/73   &57.8/72      &  56.0/73  &
55.8/73
&32.2/37   &33.1/37  & 32.6/37 &  10.7/17  \\
%%%
\tabspace
%%%
$F_{0.5-2~{\rm
keV}}$\tablenotemark{f}~($10^{-14}$~erg~cm$^{-2}$~s$^{-1}$)
     &$7.7\pm0.1$   &$7.7\pm0.1$  &$7.7\pm0.1$ & $7.7\pm0.1$
& $7.5\pm0.3$  & $7.5\pm0.3$   & $7.5\pm0.3$ & $12\pm1$   \\
%%%
\tabspace
%%%
$F_{0.5-10~{\rm
keV}}$\tablenotemark{f}~($10^{-14}$~erg~cm$^{-2}$~s$^{-1}$)
     &$8.0\pm0.5$   &$8.6\pm0.6$   & $8.1\pm0.5$ & $8.0\pm0.5$
& $7.8\pm0.8$  & $7.9\pm0.8$  & $7.8\pm0.8$ &  $16\pm2$ \\
%%%
\tabspace
%%%
$F_{0.5-2~{\rm
keV}}$\tablenotemark{g}~($10^{-13}$~erg~cm$^{-2}$~s$^{-1}$)
     &$1.29\pm0.03(\pm0.05)$ &$1.40\pm0.03(\pm0.08)$  &$1.64\pm0.03(\pm0.09)$ &$1.78\pm0.04(\pm0.09)$
& $1.4\pm0.1(\pm0.2)$  & $1.8\pm0.1(\pm 0.2)$  & $1.9\pm0.1(\pm 0.2)$ &   $2.8\pm0.2(^{+1.1}_{-0.4})$ \\
%%%
\tabspace
%%%
$F_{0.5-10~{\rm
    keV}}$\tablenotemark{g}~($10^{-13}$~erg~cm$^{-2}$~s$^{-1}$)
&$1.32\pm0.06(\pm0.05)$ &$1.49\pm0.06(\pm0.08)$
&$1.68\pm0.06(\pm0.09)$ &$1.81\pm0.04(\pm0.09)$ &
$1.4\pm0.1(\pm0.2)$ & $1.8\pm0.1(\pm0.2)$ & $2.0\pm0.1(\pm 0.2)$ &
$3.2\pm0.2(^{+1.3}_{-0.4})$ \\ \enddata \tablenotetext{a}{{\tt
NSA} model, assuming a fixed neutron star mass of 1.4\,$\rm
M_\odot$} \tablenotetext{b}{{\tt NSATMOS} model, assuming a fixed
neutron star mass of 1.4\,$\rm M_\odot$} \tablenotetext{c}{Neutron
star radius assuming a distance of 6.3~kpc}
\tablenotetext{d}{Powerlaw normalization at 1\,keV.}
\tablenotetext{e}{Lower limit constrained to $>5$.}
\tablenotetext{f}{Absorbed flux.} \tablenotetext{g}{Unabsorbed
flux.} \tablecomments{Parameter errors are 90\% confidence for one
  interesting parameter (i.e., $\Delta \chi^2=2.71$). Flux errors are
  68\% confidence (values in parantheses are systematic errors - see
  Appendix).}

\end{deluxetable}

\begin{deluxetable}{lccc}
\setlength{\tabcolsep}{0.02in} \tabletypesize{\footnotesize}
\tablewidth{0pt} \tablecaption{Best-fit {\tt NSATMOS} spectral
parameters for a joint fit of the {\it Chandra} and {\it
XMM-Newton} spectra \label{tab:tab3}}
\tablehead{\colhead{\textbf{Parameters}}  & Chandra &\multicolumn{2}{c}{XMM-Newton} \\
   & & (2005) & (2007/2008)}
\startdata ${\rm N_H}$ ($10^{22}\,{\rm cm}^{-2}$) &$0.36\pm0.01$
& $0.36\pm0.04$  &
   $0.33^{+0.02}_{-0.04}$ \\
%%%
\tabspace
%%%
$\log T_{eff}$ (log K)\tablenotemark{a} & $6.48^{+0.02}_{-0.07}$  & $6.45^{+0.02}_{-0.13}$ & $6.45^{+0.02}_{-0.20}$ \\
%%%
\tabspace
%%%
${\rm A_{PL}}$ ($10^{-5}\,\gamma/{\rm keV/cm^2/sec}$) & $1.2^{+0.3}_{-0.4}$ & \nodata  & \nodata  \\
%%%
\tabspace
%%%
$M_{NS}$ (${\rm M_\odot}$) &\multicolumn{3}{c}{$1.66^{+0.02}_{-0.58}$} \\
%%%
\tabspace
%%%
$R_{NS}$ (km) &\multicolumn{3}{c}{$5.55^{+0.01}_{-0.06}$} \\
%%%
\tabspace
%%%
$F_{0.5-10~{\rm
    keV}}$\tablenotemark{b}~($10^{-13}$~erg~cm$^{-2}$~s$^{-1}$) & $3.2\pm{0.2}(\pm0.3)$
& $2.0\pm{0.1}(\pm0.2)$ & $1.9\pm0.1(\pm0.1)$ \\
%%%
\tabspace
%%%
$\chi^2/$d.o.f.  &\multicolumn{3}{c}{99.6/128} \\ \enddata
\tablecomments{A $\Gamma\equiv2$ powerlaw was included for only
the
  {\it Chandra} spectra. The neutron star distance was fixed to
  6.3\,kpc. Parameter errors are 90\% confidence for one interesting
  parameter (i.e., $\Delta \chi^2=2.71$).  Flux errors are 68\%
  confidence (values in parantheses are systematic errors - see
  Appendix.}
\tablenotetext{a}{Upper bound of the $\log T_{eff}$ is constrained
to be $<6.5$.} \tablenotetext{b}{Unabsorbed flux.}
\end{deluxetable}

\clearpage

\end{landscape}

\appendix

\section{Flux Error Bars}

It is impossible to uniquely invert the (imperfectly known) X-ray
detector response matrices to yield a completely
model-independent, yet accurate, estimate of the detected flux for
an observed source. This also leads to difficulties in deriving
error bars for any estimate of the flux.  Here we use the fact
that the ``deconvolved flux'' is a reasonably close estimate of
the ``model flux'' to estimate the error bars on this latter
quantity as presented in Table~\ref{tab:tab2}.  Specifically, we
define the ``deconvolved photon flux'' in Pulse Height Analysis
(PHA) channel $h$ as:
\begin{equation}
F(h) ~=~ \frac{C(h)-B(h)}{T \int R(E,h) A(E) ~dE} ~~,
\label{eq:flux}
\end{equation}
where $C(h)-B(h)$ are the background subtracted counts in channel
$h$, $T$ is the observation integration time, $R(E,h)$ is the
detector response matrix, and $A(E)$ is the detector effective
area.  (See \citealt{davis:01a} for a more in depth discussion of
the meanings of these terms.)  This is the same deconvolution used
to create Fig.~\ref{fig:fig3}, and it is independent of assumed
model.  The deconvolved energy flux in a given band is then
determined by multiplying the above photon flux by the midpoint
energy of the PHA bin (determined from the {\tt EBOUNDS} array of
the response matrix), and summing over the channels within the
given energy band of interest.  The error is determined from the
sum in quadrature of the error from each non-zero bin, which in
turn is determined from the counting statistics of the source and
background.

This flux estimate can be compared to the model photon flux, which
is determined by integrating the best fit model over the energy
band of interest.  For all the models considered in this paper,
these two estimates agree to within 3\% when applied to the
\emph{absorbed} 0.5--2\,keV flux and agree to within 9\% when
applied to the \emph{absorbed} 0.5--10\,kev flux.  We therefore
use the model flux as our flux estimate; however, we use the
deconvolved flux error bars, scaled by the ratio of model to
deconvolved flux, as the estimate of our flux errors.  Note that
with this definition it is possible for the 0.5--10\,keV flux to
have a $-1\sigma$ bound below that of the 0.5--2\,keV flux.  This
might be indicative of unmodeled flux in the 3--10\,keV band;
however, we lack the statistics to describe the flux in this band
with any degree of accuracy.  From this point of view, the flux
estimate with the least amount of systematic uncertainty is the
0.5--2\,keV absorbed flux.

When determining the statistical errors on the unabsorbed flux, we
further scale the above error estimates by the energy-dependent
ratio of the unabsorbed to absorbed model flux.  The systematic
error bars on the unabsorbed fluxes are then determined
individually for each spectral model.  We freeze the fitted value
of the neutral column at its $\pm1\sigma$ limits, refit the
spectra, and then determine the fluxes as described above.  The
deviations of the unabsorbed fluxes at these two absorption limits
compared to the flux obtained when using the best fit parameter
values are assigned as the systematic error bars.  Note that such
systematic errors are for a given model, and do not reflect the
systematic errors on the unabsorbed fluxes obtained from comparing
different assumed spectral models.

For the cases of fitting the spectra with a blackbody and a
powerlaw we employ one other estimate of the flux error bars for
the individual model components.  {\tt ISIS} allows one to set any
given parameter to be an arbitrary function of any other set of
fit parameters.  It therefore was straightforward to recast the
spectral fit from dependence upon blackbody normalization and
temperature to dependence upon unabsorbed 0.5--10\,keV blackbody
flux and temperature (i.e., the blackbody normalization can be
written as a function of those two parameters).  Likewise, the
powerlaw fit parameters were recast to depend upon unabsorbed
0.5--10\,keV flux and powerlaw slope. We thus were able to use
direct fitting methods to generate the error contours shown in
Fig.~\ref{fig:fig4}.

\appendix

\section{Flux Error Bars}

It is impossible to uniquely invert the (imperfectly known) X-ray
detector response matrices to yield a completely
model-independent, yet accurate, estimate of the detected flux for
an observed source. This also leads to difficulties in deriving
error bars for any estimate of the flux.  Here we use the fact
that the ``deconvolved flux'' is a reasonably close estimate of
the ``model flux'' to estimate the error bars on this latter
quantity as presented in Table~\ref{tab:tab2}.  Specifically, we
define the ``deconvolved photon flux'' in Pulse Height Analysis
(PHA) channel $h$ as:
\begin{equation}
F(h) ~=~ \frac{C(h)-B(h)}{T \int R(E,h) A(E) ~dE} ~~,
\label{eq:flux}
\end{equation}
where $C(h)-B(h)$ are the background subtracted counts in channel
$h$, $T$ is the observation integration time, $R(E,h)$ is the
detector response matrix, and $A(E)$ is the detector effective
area.  (See \citealt{davis:01a} for a more in depth discussion of
the meanings of these terms.)  This is the same deconvolution used
to create Fig.~\ref{fig:fig3}, and it is independent of assumed
model.  The deconvolved energy flux in a given band is then
determined by multiplying the above photon flux by the midpoint
energy of the PHA bin (determined from the {\tt EBOUNDS} array of
the response matrix), and summing over the channels within the
given energy band of interest.  The error is determined from the
sum in quadrature of the error from each non-zero bin, which in
turn is determined from the counting statistics of the source and
background.

This flux estimate can be compared to the model photon flux, which
is determined by integrating the best fit model over the energy
band of interest.  For all the models considered in this paper,
these two estimates agree to within 3\% when applied to the
\emph{absorbed} 0.5--2\,keV flux and agree to within 9\% when
applied to the \emph{absorbed} 0.5--10\,kev flux.  We therefore
use the model flux as our flux estimate; however, we use the
deconvolved flux error bars, scaled by the ratio of model to
deconvolved flux, as the estimate of our flux errors.  Note that
with this definition it is possible for the 0.5--10\,keV flux to
have a $-1\sigma$ bound below that of the 0.5--2\,keV flux.  This
might be indicative of unmodeled flux in the 3--10\,keV band;
however, we lack the statistics to describe the flux in this band
with any degree of accuracy.  From this point of view, the flux
estimate with the least amount of systematic uncertainty is the
0.5--2\,keV absorbed flux.

When determining the statistical errors on the unabsorbed flux, we
further scale the above error estimates by the energy-dependent
ratio of the unabsorbed to absorbed model flux.  The systematic
error bars on the unabsorbed fluxes are then determined
individually for each spectral model.  We freeze the fitted value
of the neutral column at its $\pm1\sigma$ limits, refit the
spectra, and then determine the fluxes as described above.  The
deviations of the unabsorbed fluxes at these two absorption limits
compared to the flux obtained when using the best fit parameter
values are assigned as the systematic error bars.  Note that such
systematic errors are for a given model, and do not reflect the
systematic errors on the unabsorbed fluxes obtained from comparing
different assumed spectral models.

For the cases of fitting the spectra with a blackbody and a
powerlaw we employ one other estimate of the flux error bars for
the individual model components.  {\tt ISIS} allows one to set any
given parameter to be an arbitrary function of any other set of
fit parameters.  It therefore was straightforward to recast the
spectral fit from dependence upon blackbody normalization and
temperature to dependence upon unabsorbed 0.5--10\,keV blackbody
flux and temperature (i.e., the blackbody normalization can be
written as a function of those two parameters).  Likewise, the
powerlaw fit parameters were recast to depend upon unabsorbed
0.5--10\,keV flux and powerlaw slope. We thus were able to use
direct fitting methods to generate the error contours shown in
Fig.~\ref{fig:fig4}.

\end{document}